\documentclass[aps,final,notitlepage,oneside,twocolumn,nobibnotes,nofootinbib,superscriptaddress,
noshowpacs,centertags]{revtex4-1}

\usepackage[utf8]{inputenc}
\usepackage[english]{babel}
\usepackage{graphicx}
\usepackage{latexsym}
\usepackage{amssymb}
\usepackage{amsmath}
\usepackage{psfrag}
\usepackage{float}
\usepackage[dvips]{color}

\begin{document}

\title{Escape from a black hole with spherical warp drive}
\author{Yu.N. Eroshenko}\thanks{e-mail: eroshenko@inr.ac.ru}
\affiliation{Institute for Nuclear Research RAS, prosp. 60-letiya Oktyabrya 7a, Moscow 117312, Russia}

\date{\today}

\begin{abstract}
In this paper, a class of the warp drive (WD) type metrics is proposed in the form of spherical and plane waves or shells. In particular, these metrics can describe the passage of spherical WD through the horizon of a black hole from the inside out. In this metrics, non-singular evolution of physical fields is possible, which is demonstrated by examples of scalar, vector and fermion fields. The passage of a warp-wave through the fields is accompanied by soliton-like configurations (kinks). The limiting case of Planck-scale WD can lead to the evaporation of singularities inside black holes with the escape of particles and information into outer space, and the EPR=WD conjecture can also be proposed.
\end{abstract}

\maketitle 





\section{Introduction}

In 1994, Miguel Alcubierre has proposed a way to pass the speed limit of the theory of relativity \cite{Alc94}. To do this, it is necessary to create a curvature of space-time in the form of a bubble, in front of which space shrinks, and behind it expands. This would make it possible to reach distant points in an arbitrarily short time, faster than light could reach them in a flat Minkowski space. This idea completely fits into the framework of the general theory of relativity. Violation of the light speed limit does not occur here, because one is using the curved space-time, like for wormholes. A variation of this idea is the ``Krasnikov tube'' \cite{Kra98}. 

However, to realize a warp drive (WD), the so-called exotic matter would be required, violating energy conditions \cite{LobVis04}, \cite{SanScuVis02}. Although the appearance of exotic matter is allowed in quantum systems (Casimir energy), it is not yet clear how the exotic matter can be generated and controlled. Another difficulty is that exotic matter in the original WD version would be in a huge amount \cite{PfeFor97}. Later, this energy requirement was significantly relaxed, and in \cite{Bro99}, the required energy was reduced to several solar masses. Optimized versions of the Alcubierre bubble have also been proposed \cite{Bob01}. In the work \cite{FelHei21}, the need for exotic energy was reduced to 0.01\% of the solar mass.

An exotic substance that violates energy conditions, if it exists, can lead to a number of interesting consequences. In particular, it can support wormholes. Through the two entrances to the mouth of the wormhole, signals and matter can fly much faster than along the usual path in outer space. An entrance to the wormhole behaves like a massive object, and in \cite{FroNov93} it is indicated that one entrance of a wormhole can fall inside a black hole (BH), while the second entrance can remain outside. Thus, it becomes possible to exit the BH through the wormhole. This is possible thanks to the exotic matter supporting the wormhole. In this regard, the question arises whether it is possible to fly out of a BH in another and different way by using a WD. To the best of our knowledge, this question has not yet been clarified at the mathematical level, except for  qualitative discussion in \cite{Ell04}.  

In this paper, the ``reverse engineering'' method is used to build the WD metric in a configuration that allows it to fly out of the BH. The WD in question is a spherical layer between two radial coordinates of a Schwarzschild BH. The energy-momentum tensor of exotic matter that could realise this WD is considered. In the immediate vicinity of the horizon, the BH metric is close to the flat Rindler metric, therefore, in this limit, the spherical layer turns into a flat layer resembling flat gravitational waves, but moving at superluminal speed.

An interesting question is how the passage of warp waves affects physical systems on their way. We study this question for some particular examples of scalar, vector and fermionic fields. The obtained solutions show that the passage of a warp wave leads to a shift of the fields. Moreover, since a velocity is tending to infinity, this rearrangement of fields could occur very quickly -- almost instantly in large areas of space. We hypothesize that such a rearrangement of fields at some more fundamental level can affect the quantum correlations of spatially distant systems. If the appearance of WD type solutions is possible at the Planck energy scale, then their flow from a BH could lead to the evaporation of the singularity where Planck energies are achieved.

The article is organized as follows. In the section \ref{metr}, the metric of a BH in Kruskal-Szekeres coordinates is generalized to the case of a spherical shell with WD properties. In the section \ref{plan}, the flat limit of the introduced metric is considered and the energy-momentum tensor is calculated. The section \ref{evol} explores some particular solutions for the physical fields in the metric of a plane warp wave. The section \ref{plank} discusses Planck-scale WD.


\section{Warp drive - black hole metric}
\label{metr}

Let's first consider a Schwarzschild BH with metric
\begin{equation}
ds^2=\left(1-\frac{2M}{r}\right)dT^2-\frac{dr^2}{1-\frac{2M}{r}}-r^2d\Omega^2,
\label{shw}
\end{equation}
where $d\Omega^2=d\theta^2+\sin^2\theta d\phi^2$. It is transformed to the Kruskal-Szekeres coordinates with no features on the horizon. The only difference from the standard method is that we choose the normalization factors and the integration constant in a special way for the subsequent purposes. Turtle coordinate is 
\begin{equation}
r_*=\int\frac{dr}{1-\frac{2M}{r}}=r-2M+2M\ln\left|\frac{r}{2M}-1\right|.
\end{equation}
One introduces the light coordinates 
\begin{equation}
V=T+r_*, \quad W=T-r_*,
\end{equation}
then
\begin{equation}
V=4M\ln\left|\frac{v}{4M}\right|, \quad W=-4M\ln\left|\frac{w}{4M}\right|,
\end{equation}
and $r$ is implicitly defined from the relation
\begin{equation}
vw=16M^2e^{r/2M-1}\left(1-\frac{r}{2M}\right).
\label{vw}
\end{equation}
The  past and future horizons $r=2M$ correspond to $v=0$ and $w=0$, and the singularity $r=0$ corresponds to $vw=16M^2e^{-1}$.
Let's now introduce the new space-time coordinates
\begin{equation}
t=v+w, \quad x=v-w.
\end{equation}
In the coordinates ($t,x$), the metric (\ref{shw}) takes the form
\begin{equation}
ds^2=\frac{2M}{r}e^{1-r/2M}\left[dt^2-dx^2\right]-r^2d\Omega^2.
\label{kr1}
\end{equation}
The normalization factors and the constant were chosen so that (\ref{kr1}) is close to the Minkowski metric on the horizon $r=2M$ , i.e. there is a locally Lorentzian frame here.

Now, using reverse engineering, we generalize the metric (\ref{kr1}). Namely, we include a thin layer in the state of WD motion enclosed between two values of the radial coordinate. This generalization looks like 
\begin{eqnarray}
ds^2&=&\frac{2M}{r}e^{1-r/2M}\left[dt^2+\frac{1}{4}f(\beta t-x,r)(\beta dt-dx)^2-dx^2\right]
\nonumber
\\
&-&r^2d\Omega^2,
\label{kr2}
\end{eqnarray}
where the velocity in the ($t,x$) space can exceed the speed of light ($\beta>1$), and the function $f(u)$ is nonzero in some limited interval of its argument $u_1<u<u_2$, and $f(u)=0$ outside this interval.

\begin{figure}[t]
\includegraphics[angle=0,width=0.45\textwidth]{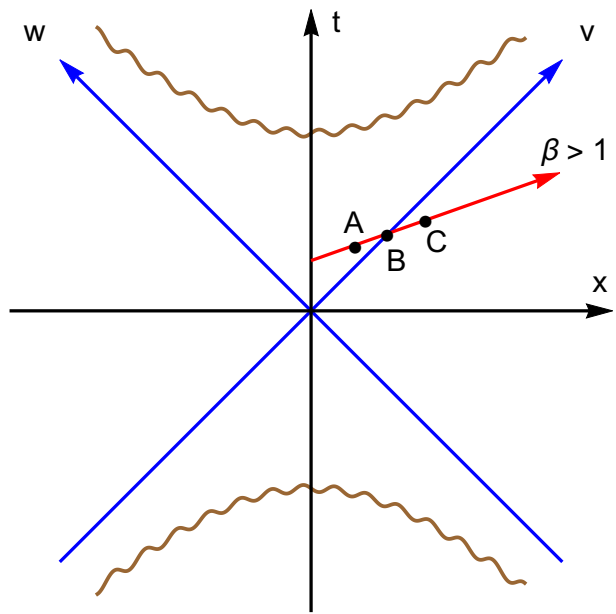}
\caption{The Kruskal-Szekeres diagram of the Schwarzschild BH. Wavy lines show singularities $r=0$. The line $\beta>1$ shows the movement of the WD region, set by the function $f(u)$ in (\ref{kr2}). The WD layer crosses the future horizon $w=0$ at point B and exits from inside the BH to the outside.}
\label{gr1}
\end{figure}
\begin{figure}[t]
\includegraphics[angle=0,width=0.45\textwidth]{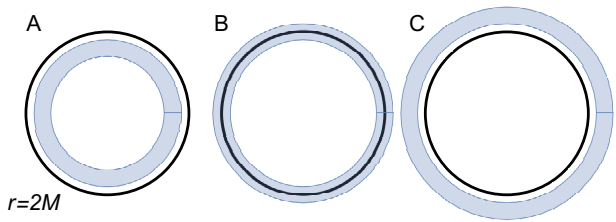}
\caption{The escape of a spherical WD from a BH. The points A, B and C correspond to the points at Fig.~\ref{gr1}.}
\label{gr2}
\end{figure}

At Fig.~\ref{gr1} the Kruskal-Szekeres diagram (of the initial BH without WD) shows the trajectory of a spherical warp layer coming out of the interior of the BH to the outside. The movement of the WD is shown by a space-like line running at an angle $\beta$ to the $x$ axis. It crosses the future horizon   $w=0$ at point B. At point A $wv>0$, therefore, according to (\ref{vw}), $r<2M$. This means that the WD is inside the BH. At point C $wv<0$, so $r>2M$. This means that the WD has flown out of the BH. And this happened as the time $t$ of the frame ($t,x$) increased. Schematically, the exit of a WD from a BH is shown at Fig.~\ref{gr2}.

Thus, formally, the reverse engineering method allows one to build a WD flying out of the BH. However, this spherical WD must be supported by exotic matter with some energy-momentum tensor, which is discussed in the next Section. Moreover, we do not consider the WD to be a test particle, so we need the full metric (\ref{kr2}) containing both the BH and WD.  The existence of physical matter capable of realizing this energy-momentum tensor remains an open question.


\section{Plane wave limit}
\label{plan}

For the general case of the metric (\ref{kr2}), the calculation of the energy-momentum tensor looks cumbersome, since $r$ is expressed implicitly in terms of coordinates. But one can introduce the following simplification. As is well known, near the horizon $r\simeq2M$, the Schwarzschild metric is close to the Rindler metric, which, in turn, is the metric of a uniformly accelerated observer, and by replacing coordinates it is transformed into the Minkowski metric. Therefore, consider a small region of space at $\theta\simeq\pi/2$ near the horizon $r\simeq2M$. The transverse coordinates are $y=2M\theta$, $z=2M\phi$, and the metric (\ref{kr2}) in this limit has the form of a flat layer
\begin{equation}
ds^2=dt^2+\frac{1}{4}f(\beta t-x,y,z)(\beta dt-dx)^2-dx^2-dy^2-dz^2.
\label{metrgen}
\end{equation}

In the case $\beta=1$ it coincides with the known metric of a plane gravitational wave \cite{Per60} (see also the problem for the Section 109 in \cite{LL-2}). But in the case $\beta\neq1$, one does not longer get a plane wave with a curvature tensor independent of $y$ and $z$. Instead, one gets a more complex flat layer. However, it must be remembered that in the limiting case of the BH metric, we consider only small $y$ and $z$.

Let's introduce the new retarded and advancing coordinates 
\begin{equation}
u=\frac{\beta t-x}2, \quad p=\frac{\beta t+x}2
\label{uv}
\end{equation}
and rewrite (\ref{metrgen}) as
\begin{equation}
ds^2=[f(u,y,z)-\eta]du^2-\eta dp^2+2(2-\eta)dudp-dy^2-dz^2,
\label{metrgen2}
\end{equation}
where denoted $\eta=1-1/\beta^2$. The coordinates (\ref{uv}) are light coordinates only in the case of $\beta=1$. There is a match
\begin{equation}
\begin{array}{lcl}
\eta\to-\infty & \mbox{for} & \beta\to 0,
\\ 
\\
\eta\to 0 & \mbox{for} & \beta\to 1,
\\ 
\\
\eta\to1 & \mbox{for} & \beta\to \infty.
\end{array}
\end{equation}
Introduction of $\eta$ plays the role of a conformal transformation that maps the infinite velocity $\beta=\infty$ to the finite value of the parameter $\eta=1$. Note that in (\ref{metrgen2}) it is possible to put formally $\eta=1$, but in the physical formulation of the problem we consider metrics with very large but finite $\beta$.  The determinant of the metric tensor
\begin{equation}
g=-[f\eta+4(1-\eta)].
\label{gdet}
\end{equation}
We assume that $f$ is not equal to zero identically in some area of the variable $u$ change, and this area will be called a warp wave. Outside the warp wave, $f(u,y,z)=0$ for any $y$ and $z$, and the metric (\ref{metrgen}) reduces to the Minkowski metric. One can consider metric (\ref{metrgen}) or (\ref{metrgen2}) even without BH as a WD moving in the Minkowski space. Therefore, we consider in this Section not only limiting case of (\ref{kr2}) but more general cases of plane warp waves.

The metric (\ref{metrgen2}) can be diagonalized as follows
\begin{eqnarray}
ds^2&=&\left[f+4\left(\frac{1}{\eta}-1\right)\right]du^2-\left[-\frac{2-\eta}{\eta^{1/2}}du+\eta^{1/2}dp\right]^2
\nonumber
\\
&-&dy^2-dz^2,
\label{tetr}
\end{eqnarray}
that is, one can introduce the tetrad
\begin{eqnarray}
e^{(0)}_\mu&=&\left(\left[f+4\left(\frac{1}{\eta}-1\right)\right]^{1/2},0,0,0\right)
\nonumber
\\
e^{(1)}_\mu&=&\left(-\frac{2-\eta}{\eta^{1/2}},\eta^{1/2},0,0\right)
\nonumber
\\
e^{(2)}_\mu&=&\left(0,0,1,0\right)
\nonumber
\\
e^{(3)}_\mu&=&\left(0,0,0,1\right).
\label{tetr2}
\end{eqnarray}
Note that this tetrad also exists in the extreme case $\eta=1$, i.e., physical fields can exist in a normal way even in an extremely fast warp wave. On the contrary, when approaching the speed of light $\eta\to0$, the system formed by this tetrad loses meaning due to the divergence of $1/\eta$.

Consider the Einstein tensor $G_{\mu\nu}=R_{\mu\nu}-g_{\mu\nu}R/2$ for the metric (\ref{metrgen2}):
\begin{eqnarray}
G_{00}&=&(2-\eta)^2\times
\nonumber
\\
&&\frac{
\left((2ff_{yy}-f_y^2+2ff_{zz}-f_z^2\right)\eta+8(f_{yy}+f_{zz})(1-\eta) 
}
{
4[f\eta+4(1-\eta)]^2
},
\nonumber
\\
G_{10}&=&-(2-\eta)\times
\nonumber
\\
&&\eta\frac{  
\left(2ff_{yy}-f_y^2+2ff_{zz}-f_z^2\right)\eta+8(f_{yy}+f_{zz})(1-\eta) 
}
{
4[f\eta+4(1-\eta)]^2
},
\nonumber
\\
G_{11}&=&\eta^2\frac{  
\left(2ff_{yy}-f_y^2+2ff_{zz}-f_z^2\right)\eta+8(f_{yy}+f_{zz})(1-\eta) 
}
{
4[f\eta+4(1-\eta)]^2
},
\nonumber
\\
G_{22}&=&\eta\frac{  
\left(2ff_{zz}-f_z^2\right)\eta+8f_{zz}(1-\eta) 
}
{
4[f\eta+4(1-\eta)]^2
},
\nonumber
\\
G_{23}&=&\eta\frac{  
\left(2ff_{yz}-f_yf_z\right)\eta+8f_{yz}(1-\eta) 
}
{
4[f\eta+4(1-\eta)]^2
},
\nonumber
\\
G_{33}&=&\eta\frac{  
\left(2ff_{yy}-f_y^2\right)\eta+8f_{yy}(1-\eta) 
}
{
4[f\eta+4(1-\eta)]^2
},
\nonumber
\end{eqnarray}
and other components are zero.

For $\eta\neq0$, the condition $\Delta f=0$ no longer gives a plane gravitational wave with curvature tensor components independent of $y$ and $z$, which was considered in \cite{Per60}. The warp wave we are considering with $\eta>0$ are localized in a flat layer characterized by the condition $f(u,y,z)\neq 0$ around the value $u=0$, but in the planes $y-z$ the curvature tensor depends on the coordinates $y$, $z$, and it is anisotropic. 

The boundary of the warp wave, where $f\to0$, requires careful consideration, since the limit values of the Einstein tensor depend on the order in which to take the limits of $\eta\to1$ and $f\to0$. In the physical formulation of the problem, $1-\eta$ can be very small, but not equal to zero. Therefore, at the warp-wave boundary, the $f\to0$ limit leads to a physically correct result. The denominator always remains $(1-\eta)^2\neq0$, and the numerator tends to zero together with the derivatives $f_y$, etc., therefore, when approaching the boundary of the warp wave, the Einstein tensor tends to zero.

Let's narrow down the class of metrics under consideration. Consider the metrics where only terms with $1-\eta$ remain in the numerators of $G_{\mu\nu}$. 
These condition leads to equations 
\begin{equation}
2ff_{yy}-f_y^2=0,
\end{equation}
\begin{equation}
2ff_{zz}-f_z^2=0,
\end{equation}
\begin{equation}
2ff_{yz}-f_yf_z=0.
\end{equation}
The positive solution of these equations has the form
\begin{equation}
f=[s(u)y+\sigma(u)z+\varepsilon(u)]^2,
\label{fgse}
\end{equation}
where $s(u)$, $\sigma(u)$ and $\varepsilon(u)$ are arbitrary functions. The only condition that we imposed earlier is that these functions are nonzero in the limited interval of the variable $u$. Therefore, one gets
\begin{eqnarray}
G_{00}&=&(2-\eta)^2
\frac{
4(s^2+\sigma^2)(1-\eta) 
}
{
[f\eta+4(1-\eta)]^2
},
\\
G_{10}&=&-(2-\eta)\eta\frac{  
4(s^2+\sigma^2)(1-\eta) 
}
{
[f\eta+4(1-\eta)]^2
},
\\
G_{11}&=&\eta^2\frac{  
4(s^2+\sigma^2)(1-\eta) 
}
{
[f\eta+4(1-\eta)]^2
},
\\
G_{22}&=&\eta\frac{  
4\sigma^2(1-\eta) 
}
{
[f\eta+4(1-\eta)]^2
},
\\
G_{23}&=&\eta\frac{  
4s\sigma(1-\eta) 
}
{
[f\eta+4(1-\eta)]^2
},
\\
G_{33}&=&\eta\frac{  
4s^2(1-\eta) 
}
{
[f\eta+4(1-\eta)]^2
}.
\end{eqnarray}

In the case $1-\eta>0$, the determinant (\ref{gdet}) does not vanish, which means that the warp metric is non-degenerate. The corresponding curvature scalar is
\begin{equation}
R=
\frac{
8\eta(1-\eta)(s^2+\sigma^2)
}
{
[f\eta+4(1-\eta)]^2
}.
\end{equation}
For $\eta>0$ one gets $R>0$. In the case of the fluid source it means $\rho+3p<0$ violating strong energy condition.


\section{Physical fields in the limiting metric}
\label{evol}

Next, we will consider an even simpler metric of the type (\ref{metrgen}), (\ref{fgse}) with the function
\begin{equation}
f=s(u)[y+\varepsilon]^2.
\label{fy}
\end{equation}
Let's consider in this metric the behaviour of physical fields, assuming these fields as probe, i.e. neglecting the back-reaction of their energy-momentum tensors on the metric. Thus, we assume that the metric (\ref{metrgen2}) is supported by some exotic matter. We will find some particular solutions characterizing the effects that occur during the passage of a warp wave.


   \subsection{Scalar field}

The Klein-Gordon equation for a scalar field with potential $V(\phi)$ has the form
\begin{equation}
\frac{1}{\sqrt{-g}}\frac{\partial}{\partial x^\mu}\left(\sqrt{-g}g^{\mu\nu}\frac{\partial\phi}{\partial x^\nu}\right)+\frac{\partial V}{\partial \phi}=0.
\label{kg1}
\end{equation}
Put, for example, $V=\lambda\phi^2/2$ and consider the field $\phi(u)$, which depends only on $u$. In the metric (\ref{metrgen}) with the function (\ref{fy}) in the limiting case $1-\eta\to0$, the equation (\ref{kg1}) takes the form
\begin{equation}
2s\frac{\partial^2\phi}{\partial u^2}-\frac{\partial s}{\partial u}\frac{\partial\phi}{\partial u}+2\lambda s^2(y+\varepsilon)^2\phi=0.
\end{equation}
The dependence of the last term on $y$ disappears and the equation becomes self-consistent at $\lambda=0$, i.e. for a massless scalar field. Let us consider this case. The resulting equation $2s\phi_{uu}-s_u\phi_u=0$ has a solution in the form of a kink
\begin{equation}
\phi(u)=C_1\int\limits_0^us^{1/2}(u')du'+C_2,
\end{equation}
where $C_1$ and $C_2$ are constants.
In this solution, in the region before the passage of the warp wave, the scalar field was homogeneous and had a certain value $\phi_1=\phi(u\to-\infty)=C_2$, and after the passage of the warp wave, the scalar field changed and took the value $\phi_2=\phi(u\to+\infty)$. Thus, the passage of a warp wave changes the values of homogeneous fields by a constant. However, the energy density of the scalar field outside the warp wave does not change, because a homogeneous scalar field with zero potential has the zero energy density.


\subsection{Vector field}
   
Take a vector field in diagonal calibration \cite{BogShi05} with the equations of motion  
\begin{equation}
\nabla^\mu\nabla_\mu A^\nu=0,
\label{urdvek1}
\end{equation}
where $\nabla^\mu$ denotes the covariant derivative. Consider the particular example $A^\mu=(0,0,\alpha(u),0)$ with one nonzero component depending only on $u$. In this case (\ref{urdvek1}) in the limit $\eta\to 1$ leads to the equation
\begin{equation}
\alpha_{uu}-\frac{1}{2}\frac{s_u}{s}\alpha_u+s\alpha=0.
\label{urdvek2}
\end{equation}
Let's try to find some specific $s(u)$ when we can get an analytical solution. To do this, we will look for a solution in the form of $s=\gamma \alpha_u$, where $\gamma=const$. In this case, we get a simple equation $\alpha_{uu}+2\gamma \alpha\alpha_u=0$ with the final solution
\begin{equation}
\alpha(u)=\frac{C_1}{C_2}\tanh\left[C_1C_2(C_3-u)\right],
\label{veksols}
\end{equation}
\begin{equation}
s(u)=C_2^2\left[C_1^2-C_2^2\alpha^2(u)\right],
\label{veksolg}
\end{equation}
where $C_{1,2,3}$ are arbitrary constants. 

The resulting function $s(u)$ has the form of a hillock and tends to zero at $u\to\pm\infty$, and the function $\alpha(u)$ has the form of a kink moving along with the warp wave. The field at $u\to\pm\infty$ tends to constant values. That is, as in the case of the scalar field discussed earlier, the warp wave simply causes a shift in the field. Note that (as in the case of an electromagnetic field) the physically measurable quantities are not $A^\mu$ itself, but the components of the tensor $F_{\mu\nu}=\partial A_\mu/\partial x^\nu-\partial A_\nu/\partial x^\mu$, which tend to zero at $u\to\pm\infty$. Despite this, the $A^\mu$ field can have physical effects if $A^\mu$ enters the phase of the wave function, as is the case of the Aharonov-Bohm effect. Thus, the passage of warp waves can almost instantly change the phases of wave functions in large areas of space.


\subsection{Fermionic field}

The Dirac equation in exotic metrics, including the Alcubierre metric, was considered in \cite{GarSab19}. We will consider this equation in the warp wave metric. The Dirac equation in a curved space-time is written as   
\begin{equation}
(i\gamma^\mu D_\mu-m)\psi=0,
\label{urd}
\end{equation}
where Dirac matrices $\gamma^\mu=e^{\phantom{00}\mu}_{(a)}\gamma^{(a)}$ are expressed in terms of ordinary Dirac matrices $\gamma^{(a)}$ in Minkowski space using the tetrad $e^{(a)}_\mu$, and the elongated derivative
\begin{equation}
D_\mu=\partial_\mu+iqA_\mu+\Gamma_\mu,
\end{equation}
where
\begin{equation}
\Gamma_\mu=\frac{1}{4}\gamma^{(a)}\gamma^{(b)}e^{\phantom{00}\nu}_{(a)}e_{(b)\nu;\mu},
\end{equation}
and $A_\mu$ is the 4-potential of the electromagnetic field. We assume the charge $q=0$ and use the tetrad (\ref{tetr2}).  
We will look for solutions of (\ref{urd}) in the form 
\begin{equation}
\psi(u,y)=
\left\|
\begin{array}{c}
a_1(u,y)+ib_1(u,y) 
\\ 
a_2(u,y)+ib_2(u,y)
\\ 
a_3(u,y)+ib_3(u,y)
\\
a_4(u,y)+ib_4(u,y)
\end{array}
\right\|.
\end{equation}
where all functions $a_i$ and $b_i$ are real. Then for these functions we obtain the system of equations
\begin{eqnarray}
F^{-1}\left(i\frac{\partial a_1}{\partial u}-\frac{\partial b_1}{\partial u}\right)+\frac{\partial a_4}{\partial y}+i\frac{\partial b_4}{\partial y}-m(a_1+ib_1)=0,
\nonumber
\\
F^{-1}\left(i\frac{\partial a_2}{\partial u}-\frac{\partial b_2}{\partial u}\right)-\frac{\partial a_3}{\partial y}-i\frac{\partial b_3}{\partial y}-m(a_2+ib_2)=0,
\nonumber
\\
F^{-1}\left(-i\frac{\partial a_3}{\partial u}+\frac{\partial b_3}{\partial u}\right)-\frac{\partial a_2}{\partial y}-i\frac{\partial b_2}{\partial y}-m(a_3+ib_3)=0,
\nonumber
\\
F^{-1}\left(-i\frac{\partial a_4}{\partial u}+\frac{\partial b_4}{\partial u}\right)+\frac{\partial a_1}{\partial y}+i\frac{\partial b_1}{\partial y}-m(a_4+ib_4)=0,
\nonumber
\end{eqnarray}
where denoted
\begin{equation}
F=\left[s(u)\left(y+\varepsilon\right)^2+4(1-\eta)\eta^{-1}\right]^{1/2}.
\end{equation}
Equating the real and imaginary parts in each equation separately, we obtain the system of eight equations for eight real functions. The solution is easy to find in the case of a massless field with $m=0$. In this case, the equations split into two independent sets corresponding to two helicities of a massless fermionic field. Next, we find a particular solution by putting $b_1=a_4=\xi(u,y)$. For the function $\xi$, we obtain the equation 
\begin{equation}
\pm F^{-1}\frac{\partial \xi}{\partial u}-\frac{\partial \xi}{\partial y}
\label{fxieq}
\end{equation}
with an upper sign. If we put $a_1=b_4=\xi$, we get the equation (\ref{fxieq}) with a lower sign. And similarly, for the second set of equations, into which the original system was separated. The equation for the characteristics of (\ref{fxieq}) has the form
\begin{equation}
\pm F(u,y)du=dy.
\label{har}
\end{equation}
Denote by $p^\pm(u,y)$ the first integrals of the equation (\ref{har}). Then the solution can be written as
\begin{equation}
\psi(u,y)=
\left\|
\begin{array}{c}
i\Psi_1+\Psi_2
\\ 
i\Psi_3+\Psi_4
\\ 
\Psi_3+i\Psi_4
\\
\Psi_1+i\Psi_2
\end{array}
\right\|.
\end{equation}
where arbitrary functions depending on the first integrals are denoted as follows: $\Psi_1=\Psi_1(p^+)$, $\Psi_2=\Psi_2(p^-)$, $\Psi_3=\Psi_3(p^-)$, $\Psi_4=\Psi_4(p^+)$.

The characteristics of the equation (\ref{fxieq}) begin in the region of the undisturbed field $u\to-\infty$, pass through the warp wave and go into the region behind the warp wave. We will find solutions for two limiting cases. In the undisturbed region before the warp wave $s(u)=0$, and the integrals of the equation (\ref{har}) have the form
\begin{equation}
p^\pm =2(1-\eta)^{1/2}\eta^{-1/2}\pm y.
\label{in1}
\end{equation}
This solution is simply a free fermionic field in Minkowski space, written in the coordinates of the warp metric. The same solution takes place after the passage of the warp wave at $u\to\infty$, but with a shift of $y$ and by a constant multiplier (see below).

In the warp wave domain, in the limiting case $1-\eta\ll s(y+\varepsilon)^2/4$, the integrals of the equation (\ref{har}) have the form
\begin{equation}
p^\pm =(y+\varepsilon)\exp\left\{\mp\int\limits_0^us^{1/2}(u')du'\right\}.
\label{in2}
\end{equation}
The general expression with limits (\ref{in1}) and (\ref{in2}) should follow from the exact solution of the equation (\ref{har}), however, this solution cannot be written in analytical form. If we assume that the warp wave occupies a limited interval of $u$ and smoothly tends to zero, and outside this interval the function $s(u)=0$, then the integral in (\ref{in2}) has a finite value. In this case, after the warp wave passes through the fermionic field, a general shift (linear transformation) of this field occurs. We see that, as in the case of scalar and vector fields, the passage of a warp wave can be accompanied by a general shift in the fields. Recall, however, that we demonstrated this conclusion only on some particular examples for the warp wave metric.


\section{Micro warp drives and evaporation of singuliarity}
\label{plank}

Let's note another aspect of the WD related to the transition from macroscopic to microscopic systems. The energy of exotic matter required to maintain the Alcubierre bubble \cite{Sho19}
\begin{equation}
E\sim-\frac{v_s^2R^2}{\varepsilon_s}\frac{c^2}{G},
\label{eal}
\end{equation}
where $v_s$ is the velocity of the bubble, $R$ is the radius of the bubble, and $\varepsilon_s$ is the thickness of the walls of the bubble. The (\ref{eal}) reaches huge values when the bubble dimensions are of the order of meters. But what happens if we consider the Alcubierre bubble with the scale of elementary particles? Assuming $R\sim\varepsilon_s\sim l$ and $v_s\sim c$, we have $E\sim lc^4/G$. The result depends on what scale of elementary interactions the $l$ scale is associated with. If $l$ is of the order of Planck length, then we immediately get that $E$ has the order of Planck energy. This may mean that on Planck scales (quantum gravity scales), warp solutions may be a typical property of particles and fields if exotic matter with negative mass arises on these scales. In particular, such microscopic WD could be born near the BH singularity, where Planck energies are reached. The departure of these microscopic WD from a BH would lead to partial evaporation of the singularity and the release of information from the BH.

In the work of L. Susskind and J. Maldacena \cite{MalSus13}, a hypothesis was put forward that quantum correlations mean that there is a connection between systems in the form of wormholes. Based on this, it was proposed that the Einstein--Podolsky--Rosen effect can be associated with Einstein-Rosen bridges, ``EPR=ER'' in short.  In this paper, we have shown that the passage of warp waves at a speed tending to infinity can change the states of physical fields. Based on this, it can be assumed that at some more fundamental level, quantum correlations can be caused by signals that have the properties of a WD and can propagate over long distances almost instantly. It would be especially interesting to find such solutions in the form of localized packages. Therefore, in addition to the ``EPR=ER'' conjecture, it can also be hypothesized that quantum correlations are supported by microscopic WD, i.e., ``EPR=WD''.


\section{Conclusion}

In modified gravity, superluminal motion is indeed possible, for example, of sound waves in the effective metric \cite{BabMukVik08}, \cite{MirRubVol21}. In the concept of a WD in general relativity, motion only looks like superluminal to an external observer (when compared with motion in Minkowski space), but in fact the WDs motion is not superluminal. Locally, their movement is always subluminal.

Within the framework of general relativity, exotic matter that violates energy conditions is necessary for the implementation of a WD, although options without violating them are also offered. Exotic matter could also be responsible for the existence of wormholes. In particular, through a wormhole, it would be possible to move signals and matter from the inside of a BH to the outside \cite{FroNov93}. In this paper, we have shown that the WD also makes it possible to fly out of the BH in the direction of increasing radius in a finite proper time of the WD. 

We considered the exit of a spherical warp wave, concentric with the horizon of a BH. But, apparently, the specific configuration of the WD does not matter in principle. Arguments were given \cite{Quora} that all directions inside the BH lead to a singularity, and that with any choice of direction, the WD will move towards the singularity. More precisely, in the $T$-region, a singularity will necessarily occur at some moment of time, regardless of the direction of movement. We think that this argument is valid only if one is moving inside or along the light cone. For a DW flying outside the light cone, there is a freedom to choose the direction from the singularity, the direction with increasing radius and the direction from the BH to the outside space.

How it is possible to set the initial state of motion of the warp wave is still unclear. It is also unknown at this stage whether solutions in the form of warp waves or some limited systems can be realised in nature.

An interesting, but still highly hypothetical possibility is the appearance of WD solutions on the Planck scale. Such microscopic WDs could be born near the BH singularity and fly out of the BH, which would give a new channel of singularity evaporation and the BH's evaporation itself. Another hypothetical possibility is the exchange of superluminal signals between spatially separated quantum objects. This exchange could be responsible for some cases of quantum entanglement between objects. This leads to the complement of the known ``EPR=ER'' conjecture \cite{MalSus13} by the ``EPR=WD'' conjecture.


\end{document}